\pdfoutput=1
\documentclass[10pt]{iopart}
\usepackage{graphicx}
\usepackage{latexsym}

\usepackage{iopams}

\def\aap{A\&A\,  }
\def\aj{AJ  }
\def\apj{ApJ\,  }

\def\sn1987a{SN \,1987A\,}
\def\s1006{SN \,1006\,}

\begin{document}
\title
{
 Three dimensional evolution of  SN 1987A
in a
self-gravitating disk.
}
\vskip  1cm
\author     {L. Zaninetti}
\address    {Dipartimento  di Fisica ,
 via P.Giuria 1,\\ I-10125 Turin,Italy }
\ead {zaninetti@ph.unito.it}

\begin {abstract}
The  introduction  of a exponential or power
law   gradient
in the interstellar medium  (ISM)
allows to produce  an asymmetric
evolution  of the supernova remnant (SNR)
when the  framework of the thin layer approximation
is adopted.
Unfortunately
both the  exponential and power law
gradients for the  ISM do  not have
a well defined  physical meaning.
The physics  conversely is well
represented  by an isothermal self-gravitating disk
of particles whose
velocity is everywhere Maxwellian.  .
We derived a law of motion
in the framework of   the thin      layer approximation
with  a control parameter  of the swept mass.
The photon's losses  ,that are often neglected
in the thin layer approximation,
are modeled   trough a velocity  dependence.
The developed  framework is applied  to
SNR 1987A and  the   three observed rings
are simulated.
\end{abstract}
\vspace{2pc}

\maketitle

\section{Introduction}
The expansion of
supernova remnant (SNR)
is usually  explained  by
 the theory  of the shock waves,
as an example
\cite{Chevalier1982a,Chevalier1982b}
analyzes  self-similar solutions
with varying inverse power law exponents
for the density profile of the advancing matter,
$R^{-n}$, and ambient  medium,
$R^{-s}$.
The SNRs conversely often present a
weak departure from the circular symmetry,
as an example in
\s1006   a ratio of 1.2
between maximum and minimum radius
has been measured, see \cite {Reynolds1986}
or
great asymmetry
as   the triple  ring  of \sn1987a , see
\cite{Morris2007,Tziamtzis2011}.

The asymmetries of SNRs are hardly  explained
by the shock theory  which is
usually  developed in spherical or
planar symmetry.
The presence of gradients in the density of the medium
allows to explain such asymmetries
when the conservation of the momentum is adopted.
As an example  an exponential decay allows to
model \s1006  and   \sn1987a, see  \cite{Zaninetti2012b}.
At the same time the various  gradients
that can be adopted  such as Gaussian,power law
and exponential  do not have a precise physical basis.
The auto-gravitating  Maxwellian medium
provides a gradient given by the secant law.
In this  paper we will answer the following questions.
\begin {itemize}
\item
Is it possible to deduce an analytical formula for
the temporal evolution of an SN  in the presence
of a vertical profile density as given by an ISD?

\item
Is it possible to deduce  an   equation
of motion for  SN   with an adjustable
parameter  that can be found from  a numerical  analysis
of the diameter--time relationship?

\item
Is it possible to simulate the appearance of  the triple
ring system of \sn1987a  ?

\end{itemize}

\section{A modified  thin layer approximation}

\label{sec_motion}

The thin layer approximation
with non cubic dependence (NCD) , $p$ ,
in classical physics
assumes that only a fraction
of the total mass enclosed in the  volume
of the expansion
accumulates  in a thin shell just after
the shock front.
The  global mass between
$0$ and $R_0$ is    $\frac {4}{3} \pi
\rho  R_0^3  $ where $\rho$ is the density
of the ambient medium.
The swept mass included in the  thin layer
which characterizes the
expansion is
\begin{equation}
M_0 =( \frac {4}{3} \pi \rho  R_0^3)^{\frac{1}{p}}
\quad  .
\end{equation}
The  mass swept between  $0$ and $R$
is
\begin{equation}
M =( \frac {4}{3} \pi \rho  R^3)^{\frac{1}{p}}
\quad  .
\end{equation}
The conservation of radial momentum requires that,
after the initial  radius $R_0$,
\begin{equation}
M   V =
M_0 V_0
\quad ,
\end{equation}
where $R$ and $V$   are  the
radius and velocity
of the advancing shock.
The velocity  is
\begin{equation}
\label{velocityclassicalm}
V=  V_0 (\frac {M_0}{M})^{\frac{1}{p}}
\quad .
\end{equation}
The velocity as a function of the radius
when the expansion starts at the origin
of the coordinates is:
\begin{equation}
\label{velocityclassical}
V=  V_0 (\frac {R_0}{R})^{\frac{3}{p}}
\quad .
\end{equation}
The velocity can be expressed  as
\begin{equation}
V= \frac{dR}{dt}
\quad  ,
\end{equation}
and the integration  between $R$ , $t$  and $R_0$ , $t_0$
gives
\begin{equation}
p{R}^{1+3\,{p}^{-1}} \left( 3+p \right) ^{-1}-p{R_{{0}}}^{1+3\,{p}^{-1
}} \left( 3+p \right) ^{-1}=V_{{0}}{R_{{0}}}^{3\,{p}^{-1}} \left( t-t_
{{0}} \right)
\quad .
\end{equation}
We solve the previous equation  for $R$ and we obtain
\begin{eqnarray}
R(t) = \nonumber  \\
\left( {R_{{0}}}^{1+3\,{p}^{-1}}+ \left( 3+p \right) V_{{0}}{R_{{0}}}
^{3\,{p}^{-1}} \left( t-t_{{0}} \right) {p}^{-1} \right) ^{{\frac {p}{
3+p}}}
\quad .
\label{rtclassical}
\end{eqnarray}
The velocity
is:
\begin{eqnarray}
V  =  \frac{N}{D}                                 \\
where                             \nonumber  \\
N=
                                  \nonumber  \\
( {{\it R_0}}^{1+3\,{p}^{-1}}+ ( 3+p  ) \,  \nonumber \\
\,  {\it V_0}\,{{
\it R_0}}^{3\,{p}^{-1}} ( t-{\it t_0}  ) {p}^{-1}  ) ^{{
\frac {p}{3+p}}}{\it V_0}\,{{\it R_0}}^{3\,{p}^{-1}}p
\nonumber \\
D={{\it R_0}}^{1+3\,{p}^{-1}}p+3\,{\it V_0}\,{{\it R_0}}^{3\,{p}^{-1}}t+{
\it V_0}\,{{\it R_0}}^{3\,{p}^{-1}}tp
\nonumber  \\
-3\,{\it V_0}\,{{\it R_0}}^{3\,{p}^{-
1}}{\it t_0}-{\it V_0}\,{{\it R_0}}^{3\,{p}^{-1}}{\it t_0}\,p
\, .
                        \nonumber \\
\label{velocitym}
\nonumber
\end{eqnarray}
Equation (\ref{rtclassical}) can  also be solved
with a  solution of type
$R=K(t-t_0)^{\alpha}$,
$k$ being a constant,
and the asymptotic  result is:
\begin{equation}
\label{radiussimilar}
R(t) =
 \left( \frac{  \left( 3+p \right) V_{{0}}{R_{{0}}}^{3/p}
\left( t-t_
{{0}} \right) }{p}  \right) ^{{\frac {p}{3+p}}}
\quad .
\end{equation}
The  asymptotic   solution  for the  velocity is
\begin {eqnarray}
V(t) =       \nonumber  \\
 \left( 3+p \right) ^{-\frac {3} { 3+p}}{p}
^{\frac {3}{3+p}}
{{\it V_0}}^{{\frac {p}{3+p}}}{{\it R_0}}
^{\frac{3}{3+p}}
\left( t-{\it t_0} \right) ^{-\frac {3} { 3+p}}
\quad  .
\label{velocitysimilar}
\end{eqnarray}

\section{Asymmetrical  law  of motion with NCD}

\label{sec_asymmetry}

Given the Cartesian   coordinate system
$(x,y,z)$,
the plane $z=0$ will be called the equatorial plane,
$z= R \sin ( \theta) $,
where $\theta$ is the latitude angle
which has range
$[-90 ^{\circ}  \leftrightarrow  +90 ^{\circ} ]$,
and   $R$ is the distance from the origin.
In our  framework,  the polar angle of the spherical
coordinate system
is  $90 - \theta$.
Here we adopt
the density profile of a thin
self-gravitating disk of gas which is characterized by a
Maxwellian distribution in velocity and  distribution which varies
only in the z-direction (ISD).
The  number density
distribution is
\begin{equation}
n(z) = n_0 sech^2 (\frac{z}{2\,h})
\quad  ,
\label{sech2}
\end{equation}
where $n_0$ is the density at $z=0$,
$h$ is a scaling parameter
and  $sech$ is the hyperbolic secant
(\cite{Bertin2000,Padmanabhan_III_2002}).
On assuming that the expansion starts  at
$z\approx 0$ , we can write $z=R \sin (\theta)$
and therefore
\begin{equation}
n(R,\theta) = n_0 sech^2 (\frac{R \sin (\theta) }{2\,h})
\quad  ,
\label{sech2rtheta}
\end{equation}
where  $R$ is the radius of the advancing shell.

The 3D expansion that starts at the origin
of the coordinates
will be characterized by the following
properties
\begin {itemize}
\item Dependence of the momentary radius of the shell
      on  the polar angle $\theta$ that has a range
      $[-90 ^{\circ}  \leftrightarrow  +90 ^{\circ} ]$.

\item Independence of the momentary radius of the shell
      from  $\phi$ , the azimuthal  angle  in the x-y  plane,
      that has a range
      $[0 ^{\circ}  \leftrightarrow  360 ^{\circ} ]$.
\end {itemize}
The mass swept, $M$,  along the solid angle
$ \Delta\;\Omega $,  between 0 and $R$ is
\begin{equation}
M(R)=
\frac { \Delta\;\Omega } {3}  m_H n_0 I_m(R)
+ \frac{4}{3} \pi R_0^3 n_0 m_H
\quad  ,
\end {equation}
where
\begin{equation}
I_m(R)  = \int_{R_0} ^R r^2
sech^2 (\frac{r \sin (\theta) }{2\,h})
dr
\quad ,
\end{equation}
where $R_0$ is the initial radius
and   $m_H$ the mass of the hydrogen.
The integral is
\begin{eqnarray}
I_m(R)  =  \nonumber \\
-4\,h{r}^{2} ( \sin ( \theta )  ) ^{-1} ( 1
+{{\rm e}^{{\frac {r\sin ( \theta ) }{h}}}} ) ^{-1}+4
\,{\frac {h{r}^{2}}{\sin ( \theta ) }}
\nonumber \\
-8\,{h}^{2}r\ln
 ( 1+{{\rm e}^{{\frac {r\sin ( \theta ) }{h}}}}
 )  ( \sin ( \theta )  ) ^{-2}
\\
-8\,{h}^{3}{
\it
\mathrm{Li}_{2}\/}
(-{{\rm e}^{{\frac {r\sin ( \theta ) }
{h}}}} )  ( \sin ( \theta )  ) ^{-3}
\quad ,
\end{eqnarray}
where
\begin{equation}
\mathop{\mathrm{Li}_{2}\/}\nolimits\!\left(z\right)=\sum
_{{n=1}}^{\infty}\frac{z^{n}}{n^{2}}
\quad ,
\end{equation}
is the  dilogarithm,
see  \cite{Hill1828,Lewin1981,NIST2010}.

The conservation of the momentum gives
\begin{equation}
(M(R))^{\frac{1}{p}}    \dot {R}=
(M(R_0))^{\frac{1}{p}}  \dot {R_0}
\quad  ,
\end{equation}
where $\dot {R}$  is the  velocity
at $R$ and
$\dot {R_0}$  is the  initial velocity at $R=R_0$
where  $p$   is the NCD parameter.
This  means  that only a fraction
of the total mass enclosed in the  volume
of the expansion
accumulates in a thin shell just after
the shock front.
The emission of photons
produces losses
in the momentum carried
by the advancing shell here assumed
to be proportional to $R^2  \dot {R}$.
The  generalized
conservation law is now
\begin{equation}
(M(R))^{\frac{1}{p}}    \dot {R}
+C_l R^2  \dot {R}
 =
(M(R_0))^{\frac{1}{p}}  \dot {R_0}
\quad  ,
\label{momentumlosses}
\end{equation}
where
$C_l$ is  a constant.
At  the  moment  of writing
is not possible  to  deduce the  parameter $C_l$
which controls  the quantity of momentum
carried away  by the photons due to the presence
of the  parameter $p$ which controls
the  quantity of swept mass during the expansion.
According to the previous generalized
equation (\ref{momentumlosses})
an analytical  expression  for  $V(R)$  can be found
but it is complex and therefore we omit it.
In this differential equation of the first order in $R$ the
variable can be separated and the integration
term by term gives
\begin{equation}
\int_{R_0}^{R} \left [ ( M(R))^{1/p}
+C_l R^2  \right ]  dr =
(M(R_0))^{1/p} \dot {R_0} \, ( t-t_0)
\quad  ,
\end{equation}
where  $t$ is the time and
$t_0$ the time at $R_0$.
We therefore have  an equation of the type
\begin{equation}
{\mathcal{F}}(R,R_0,h)_{NL} =
{3}^{-{p}^{-1}}{{\it R_0}}^{3\,{p}^{-1}}{\it V_0}\,
\left( t-{\it t_0}
 \right)
\quad  ,
\label{fundamental}
\end{equation}
where  $  {\mathcal{F}}(R,R_0,h)_{NL}  $ has an analytical
but complex form.
In  our case in order
to find  the root of the nonlinear
equation (\ref{fundamental})
the FORTRAN SUBROUTINE  ZRIDDR from \cite{press} has been used.

The physical units have not yet been specified, $pc$
for length and
$yr$ for time are perhaps an acceptable
astrophysical choice.
On adopting
the  previous  units, the initial velocity $V_{{0}}$ is
expressed in
$\frac{pc}{yr}$ and should be converted
into $\frac{km}{s}$; this
means that $V_{{0}} =1.020738\, 10^{-6} V_{{1}}$
where  $V_{{1}}$ is
the initial velocity expressed in $\frac{km}{s}$.

\section{Application to SN 1987A }

\subsection{The simulation}
\begin{figure}
\includegraphics[width=6cm]{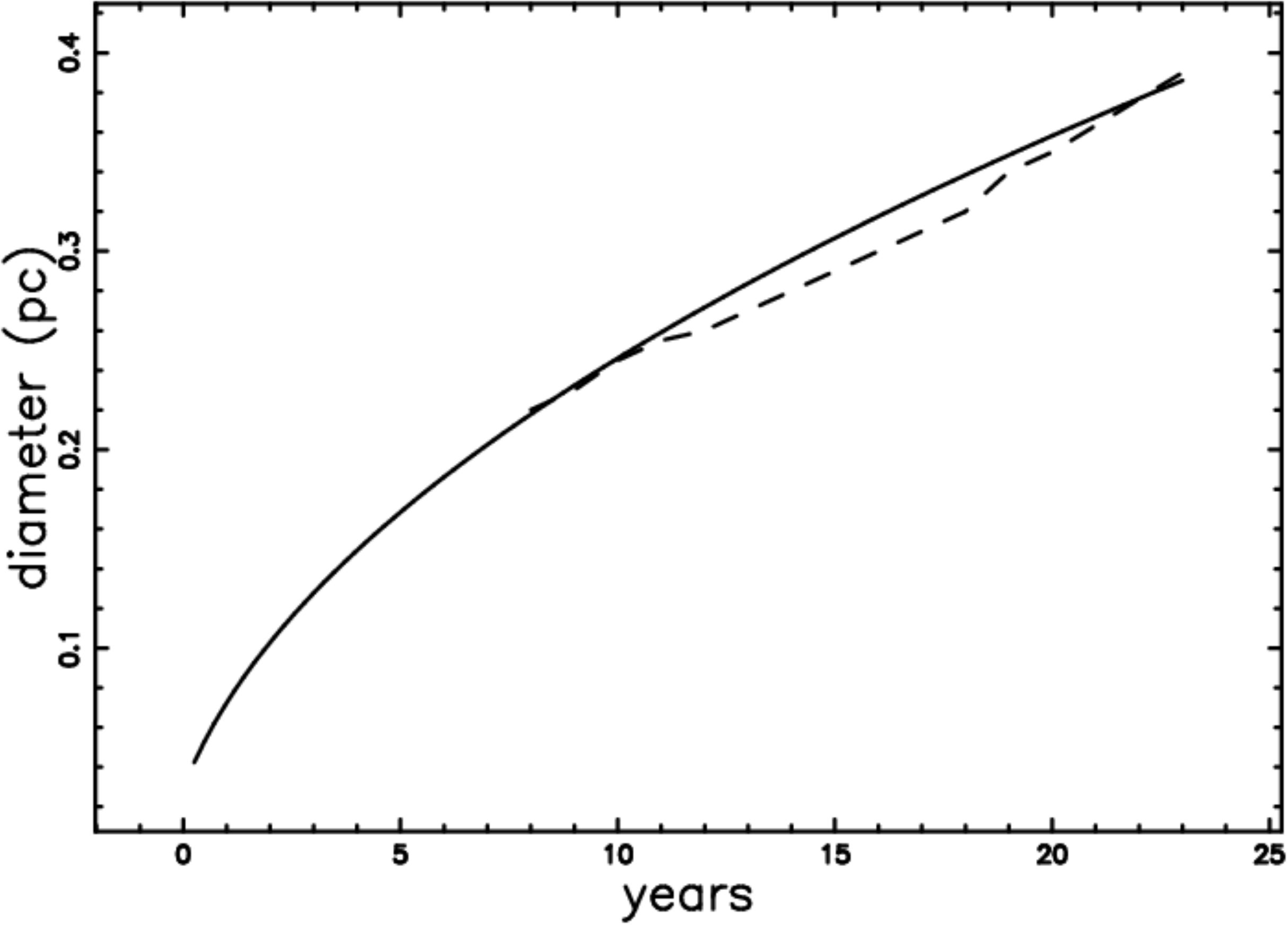}
\caption {
Diameter  as a function of time for
a self-gravitating  medium (full line)
when $\theta=0.001$ (equatorial plane)
and  astronomical data of torus only  as  extracted
from  Figure 3 in  \cite{Chiad2012} (dashed line) .
Physical parameters as in Table \ref{datafit1987a}.
          }%
    \label{auto_diametro_1987a}
    \end{figure}

\begin{figure}
\includegraphics[width=6cm]{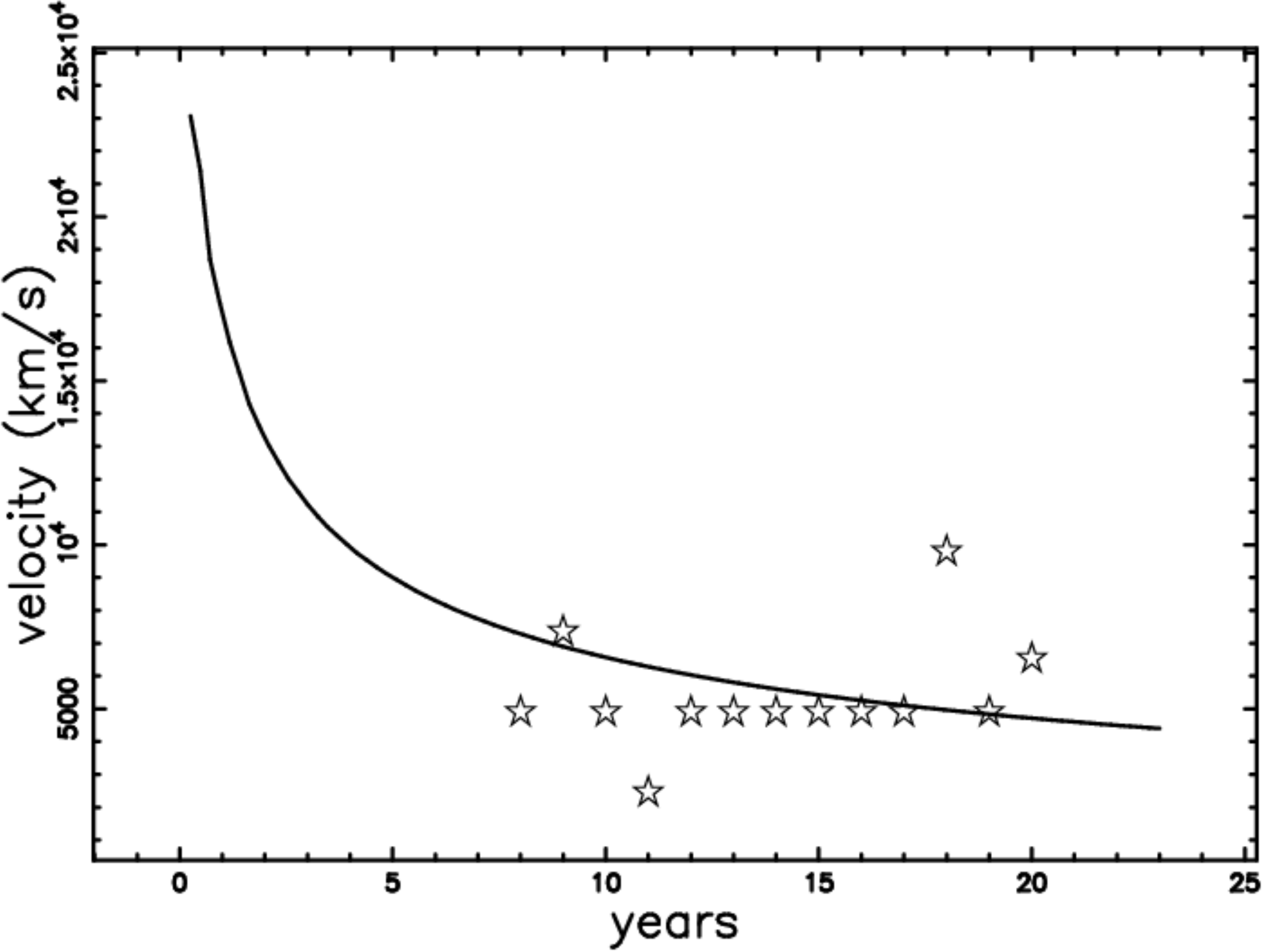}
\caption {
Velocity as a function of time for
a self-gravitating  medium (full line)
when $\theta=0.001$ (equatorial plane)
and  astronomical data of torus only  as  extracted
from  Figure 3 in  \cite{Chiad2012}  (empty stars).
Physical parameters as in Table \ref{datafit1987a}.
          }%
    \label{auto_velocita_1987a}
    \end{figure}

\begin{figure}
\includegraphics[width=6cm]{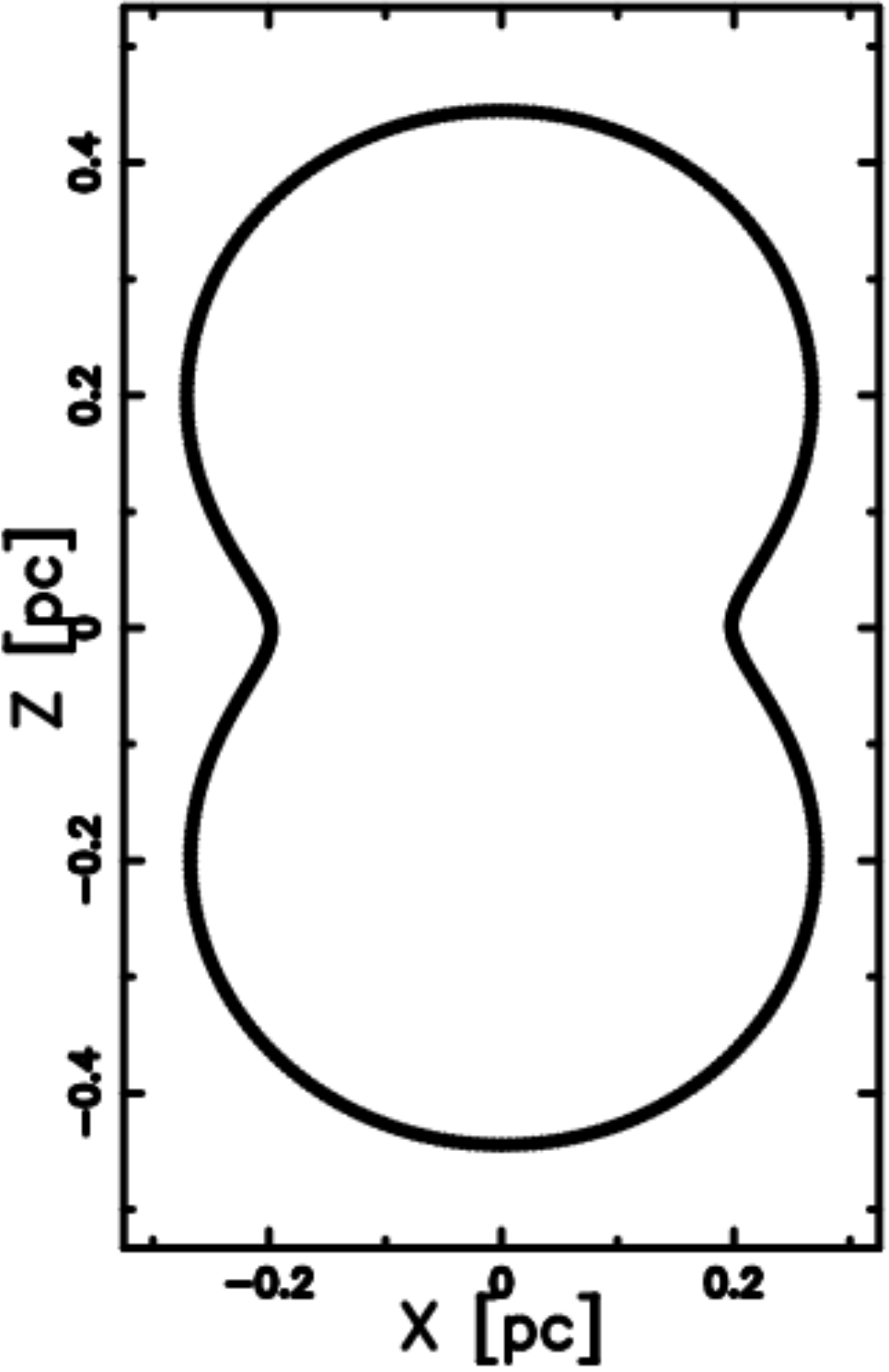}
\caption {
Section of  \sn1987a  in  the
{\it x-z}  plane.
The horizontal and vertical axis are in $pc$.
Physical parameters
as in Table \ref{datafit1987a}.
          }%
    \label{auto_cut_1987a}
    \end{figure}

\begin{figure}
\includegraphics[width=6cm]{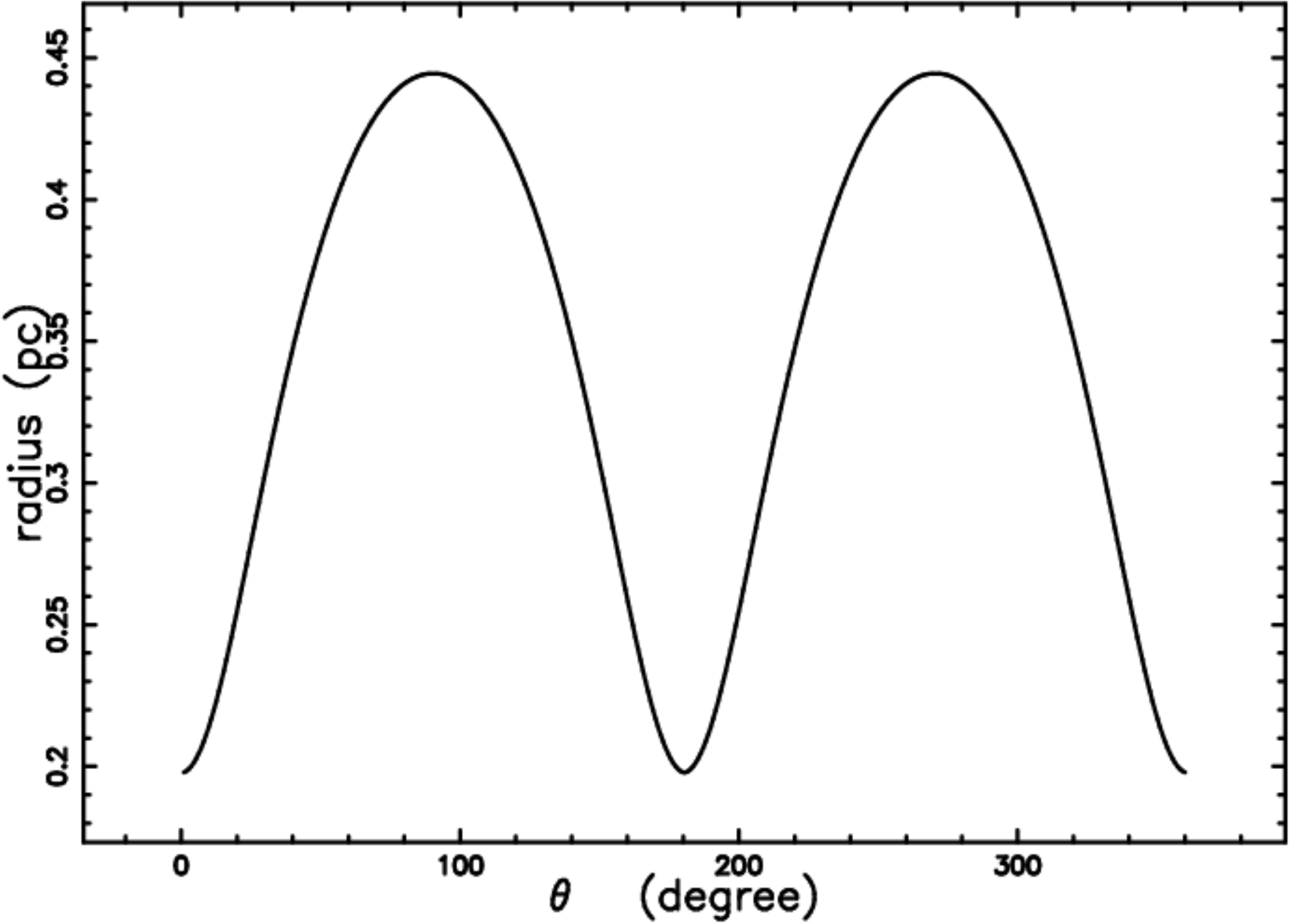}
\caption {
Radius in $pc$  of \sn1987a as a function of
the position   angle   in degrees for
a self-gravitating  medium.
Physical parameters as in Table \ref{datafit1987a}.
          }%
    \label{auto_radius_1987a}
    \end{figure}

\begin{figure}
\includegraphics[width=6cm]{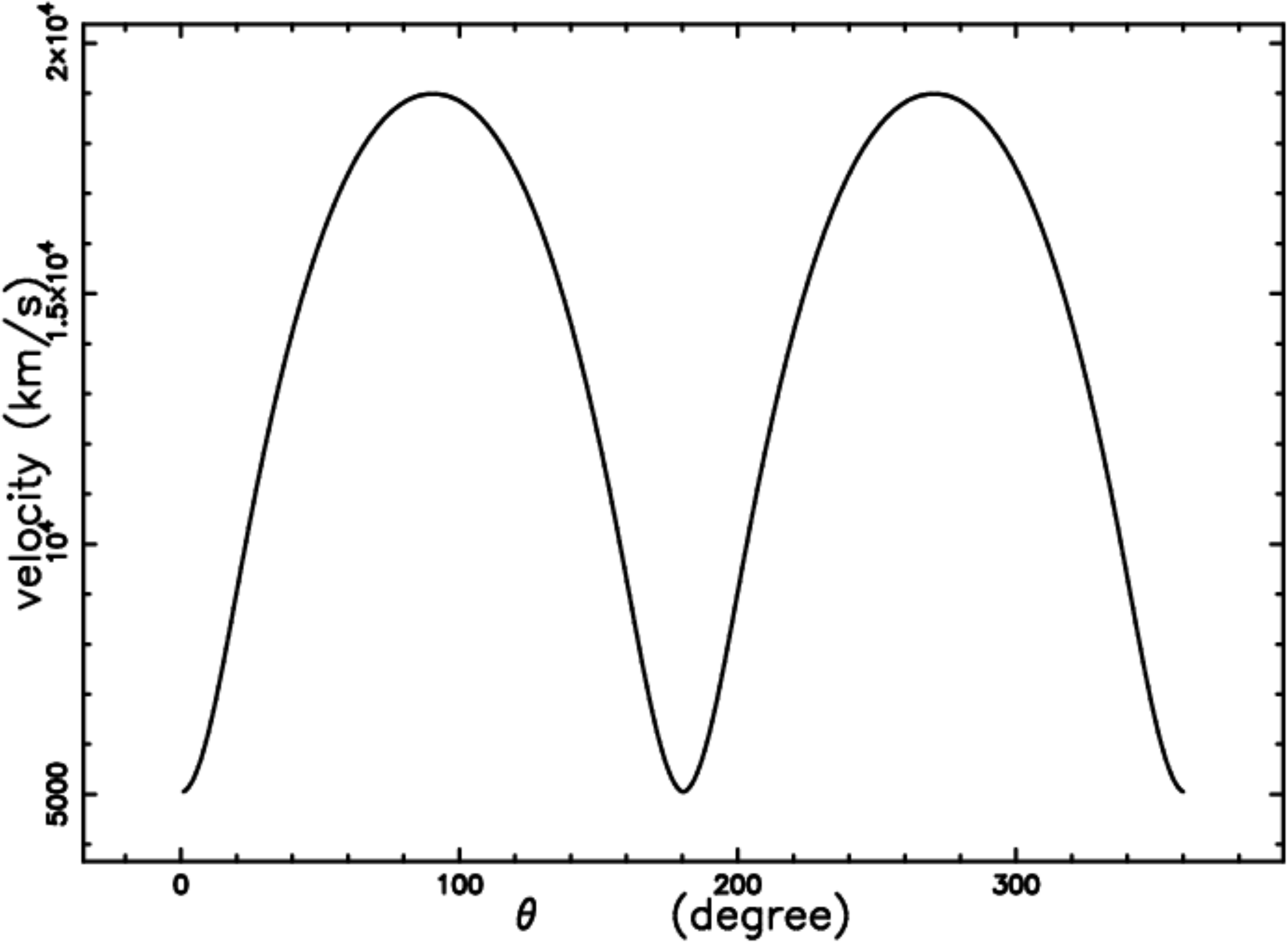}
\caption {
Velocity  in $\frac{km}{s}$  of \sn1987a as a
function of the position   angle in degrees
for a self-gravitating  medium.
Physical parameters as
in Table \ref{datafit1987a}.
          }%
    \label{auto_velocity_1987a}
    \end{figure}

The \sn1987a exploded in the Large Magellanic Cloud
in 1987.
The observed image is complex   and we will
follow the  nomenclature of \cite{Racusin2009}
which distinguish  between
torus only , torus +2 lobes and torus + 4 lobes.
In particular we  concentrate  on the torus
which is characterized  by a distance
from the center of the tube
and the radius of the tube.
Table 2 of \cite{Racusin2009} reports  the relationship
between  distance  of the torus in arcsec and  time
since  the explosion in days.
Figure 3 in \cite{Chiad2012} reports  diameter in $pc$
versus  year when the counting pixels is adopted.
Table \ref{datafit1987a}  reports the data
used in the simulation ,
Figures  \ref{auto_diametro_1987a}
and      \ref{auto_velocita_1987a}
report the diameter and the velocity in the equatorial
plane  respectively as well the observational
data as found by the counting pixels method ,
see \cite{Chiad2012}.
\begin{table}
\caption
{
Numerical value of the parameters
of the simulation  for  \sn1987a
}
\label{datafit1987a}
 \[
 \begin{array}{lcc}
 \hline
 \hline
Quantity     & Unit         &   value \\
R_0          & pc           &   0.015 \\
\dot {R}_0   & \frac{km}{s} &   26000 \\
p            & number       &   4     \\
C_l          & pc^{-2}      &   1     \\
h            & pc           &   0.011 \\
t_{0}        & yr           &   0.023 \\
t            & yr           &   23    \\
\noalign{\smallskip}
 \hline
 \hline
 \end{array}
 \]
 \end {table}
Fig. \ref{auto_cut_1987a}
reports the asymmetric expansion in
a  section crossing the center,
Figures  \ref{auto_radius_1987a}
and  \ref{auto_velocity_1987a}
report the radius  and the velocity
respectively as a function of the position
angle $\theta$.

\begin{figure}
\includegraphics[width=6cm]{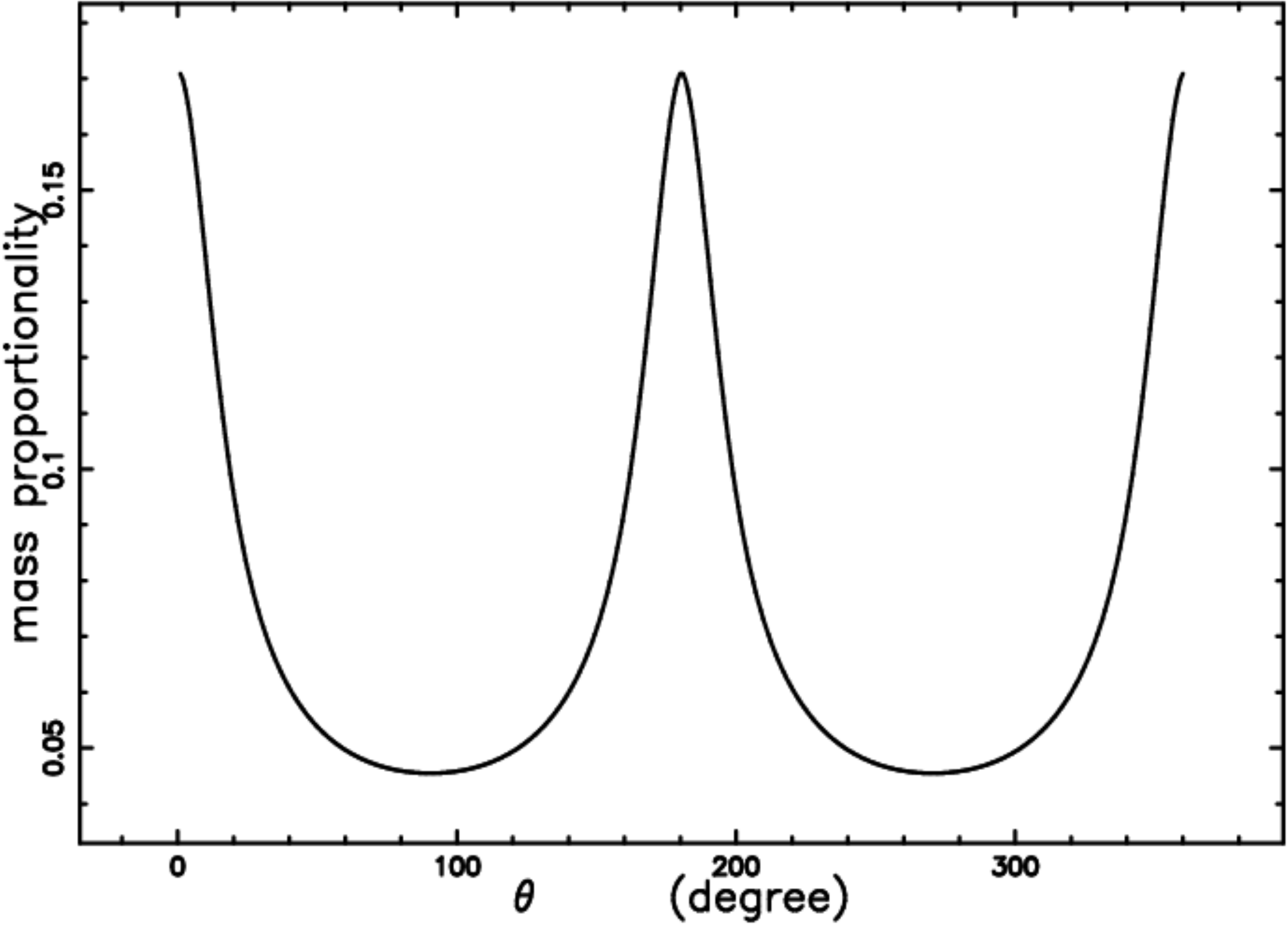}
\caption {
Swept mass   of \sn1987a
in arbitrary units
as a function of the position angle
in degrees ,
for
a self-gravitating  medium.
Physical parameters as
in Table \ref{datafit1987a}.
          }%
    \label{auto_massa_1987a}
    \end{figure}

The swept  mass  in our model is not
constant but varies with the value
of latitude , see Figure \ref{auto_massa_1987a}.
\subsection{The image}
We assume that the
emissivity is proportional to the number density.
\begin{equation}
j_{\nu} \zeta =K  C(s) \quad  ,
\end{equation}
where
$j_{\nu}$ is the emission coefficient
and  $K$ is a  constant function.
This can be the case of
synchrotron radiation in presence of an
isotropic distribution of
electrons with a power law distribution in energy, $N(E)$,
\begin{equation}
N(E)dE = K_s E^{-\gamma_f} \label{spectrum} \quad  ,
\end{equation}
where $K_s$ is a constant.
The source of synchrotron luminosity is assumed here to be
the flux of kinetic energy, $L_m$,
\begin{equation}
L_m = \frac{1}{2}\rho A  V^3
\quad,
\label{fluxkineticenergy}
\end{equation}
where $A$ is the considered area, see formula (A28)
in \cite{deyoung}.
In our  case $A=4\pi R^2$,
which means
\begin{equation}
L_m = \frac{1}{2}\rho 4\pi R^2 V^3
\quad ,
\label{fluxkinetic}
\end{equation}
where $R$  is the instantaneous radius of the SNR and
$\rho$  is the density in the advancing layer
in which the synchrotron emission takes place.
The  total observed luminosity can  be expressed as
\begin{equation}
L_{tot} = \epsilon  L_{m}
\label{luminosity}
\quad  ,
\end{equation}
where  $\epsilon$  is  a constant  of conversion
from  the mechanical luminosity   to  the
total observed luminosity in synchrotron emission.

The numerical algorithm which allows us to
build  a complex  image is now
outlined.
\begin{itemize}
\item An empty (value=0)
memory grid  ${\mathcal {M}} (i,j,k)$ which  contains
$NDIM^3$ pixels is considered
\item We  first  generate an
internal 3D surface by rotating the ideal image
 $180^{\circ}$
around the polar direction and a second  external  surface at a
fixed distance $\Delta R$ from the first surface. As an example,
we fixed $\Delta R = R/12 $, where $R$ is the
momentary  radius of expansion.
The points on
the memory grid which lie between the internal and external
surfaces are memorized on
${\mathcal {M}} (i,j,k)$ with a variable integer
number   according to formula
(\ref{fluxkinetic})  and   density $\rho$ proportional
to the swept    mass, see  Fig. \ref{auto_massa_1987a}.
\item Each point of
${\mathcal {M}} (i,j,k)$  has spatial coordinates $x,y,z$ which  can be
represented by the following $1 \times 3$  matrix, $A$,
\begin{equation}
A=
 \left[ \begin {array}{c} x \\\noalign{\medskip}y\\\noalign{\medskip}{
\it z}\end {array} \right]
\quad  .
\end{equation}
The orientation  of the object is characterized by
 the
Euler angles $(\Phi, \Theta, \Psi)$
and  therefore  by a total
 $3 \times 3$  rotation matrix,
$E$, see \cite{Goldstein2002}.
The matrix point  is
represented by the following $1 \times 3$  matrix, $B$,
\begin{equation}
B = E \cdot A
\quad .
\end{equation}
\item
The intensity map is obtained by summing the points of the
rotated images
along a particular direction.
\item
The effect of the  insertion of a threshold intensity
, $I_{tr}$ ,
given by the observational techniques ,
is now analyzed.
The threshold intensity can be
parametrized  to  $I_{max}$,
the maximum  value  of intensity
characterizing the map.
\end{itemize}

\begin{figure}
\includegraphics[width=6cm]{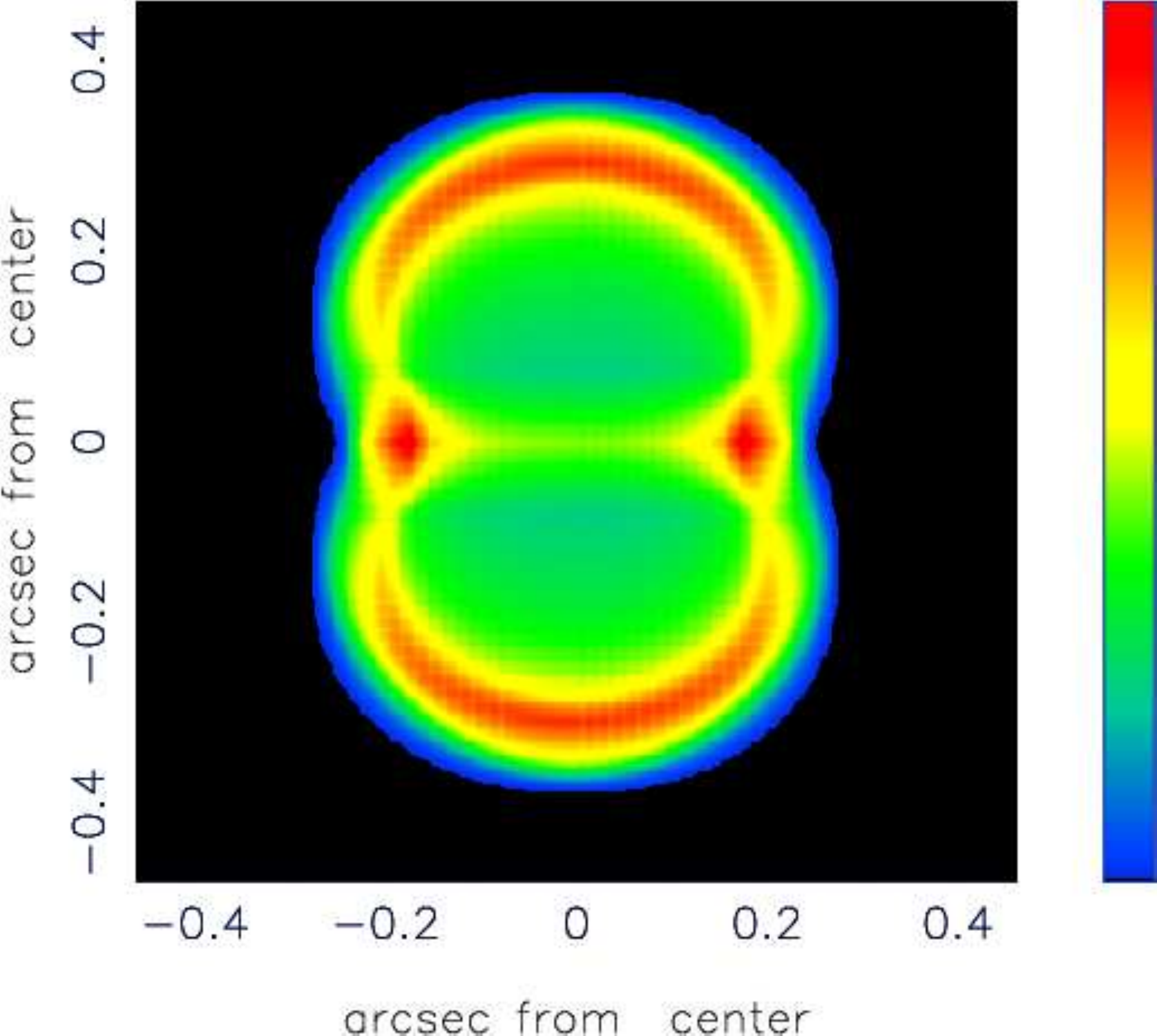}
\caption {
Map of the theoretical intensity  of
\sn1987a
in the presence of a self-gravitating  medium..
Physical parameters as in Table \ref{datafit1987a}.
The three Euler angles
characterizing the   orientation
  are $ \Phi $=180$^{\circ }$,
$ \Theta     $=90 $^{\circ }$
and   $ \Psi $=0  $^{\circ }$.
This  combination of Euler angles corresponds
to the rotated image with the polar axis along the
z-axis.}%
    \label{auto_1987a_heat}
    \end{figure}

\begin{figure}
\includegraphics[width=6cm]{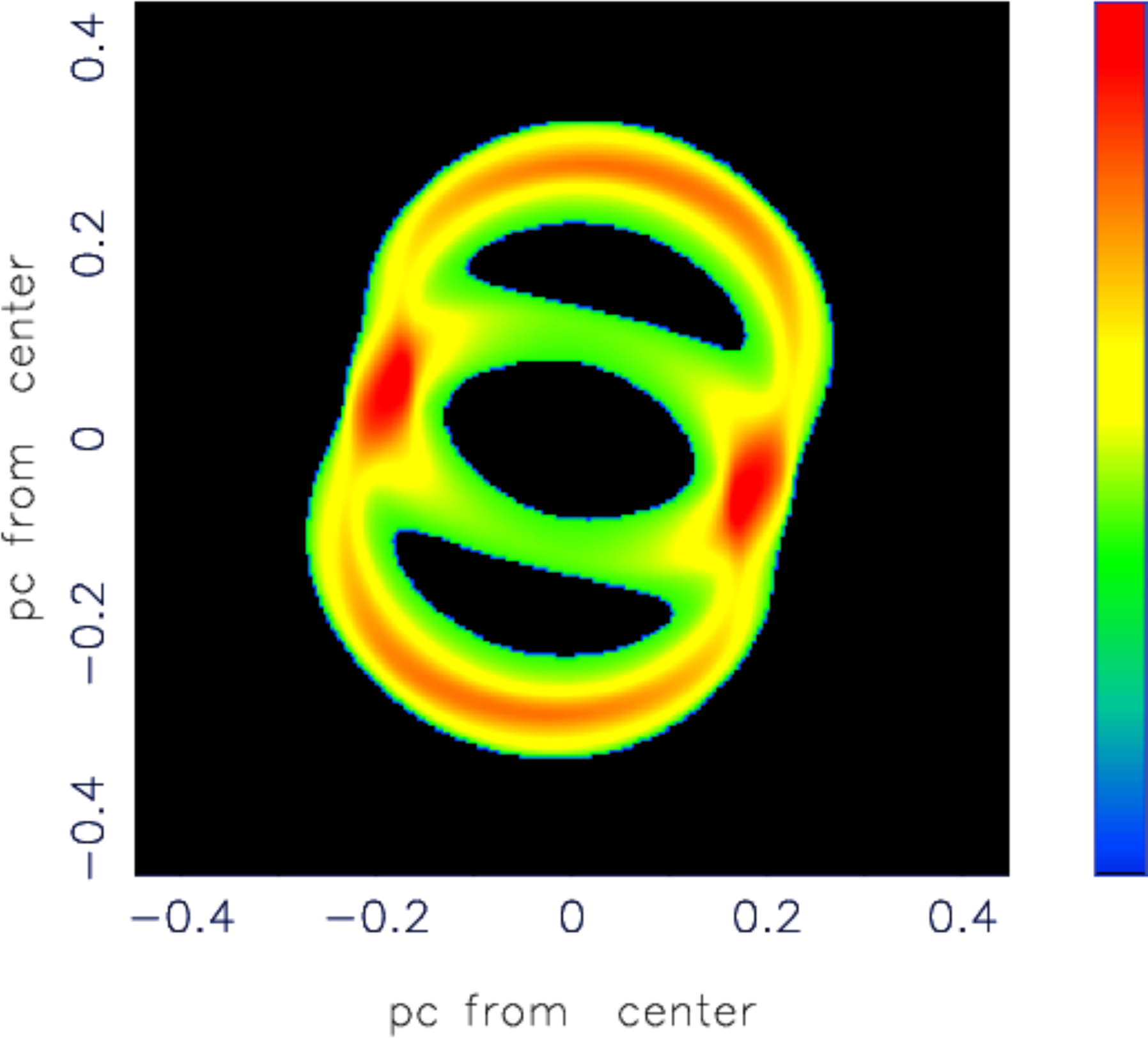}
\caption {
Model map of \sn1987a rotated in
accordance with the observations,
for a  self-gravitating  medium.
Physical parameters as in Table \ref{datafit1987a}.
the orientation of the observer
are
     $ \Phi   $=105$^{\circ }$,
     $ \Theta $=55 $^{\circ }$
and  $ \Psi   $=-165 $^{\circ }$.
This  combination of Euler angles corresponds
to the observed image.
In this map $I_{tr}= I_{max}/4$
          }%
    \label{auto_sn1987a_hole}
    \end{figure}

The image of  \sn1987a
having  polar axis along the z-direction,
is shown in Fig. \ref{auto_1987a_heat}.
The image  for a realistically rotated \sn1987a
is shown   in Fig. \ref{auto_sn1987a_hole}
and a comparison should be done
with  the observed triple ring system as reported
in Figure 1 of \cite{Tziamtzis2011}
where the outer lobes are explained by light-echo
effects.

\begin{figure*}
\begin{center}
\includegraphics[width=10cm]{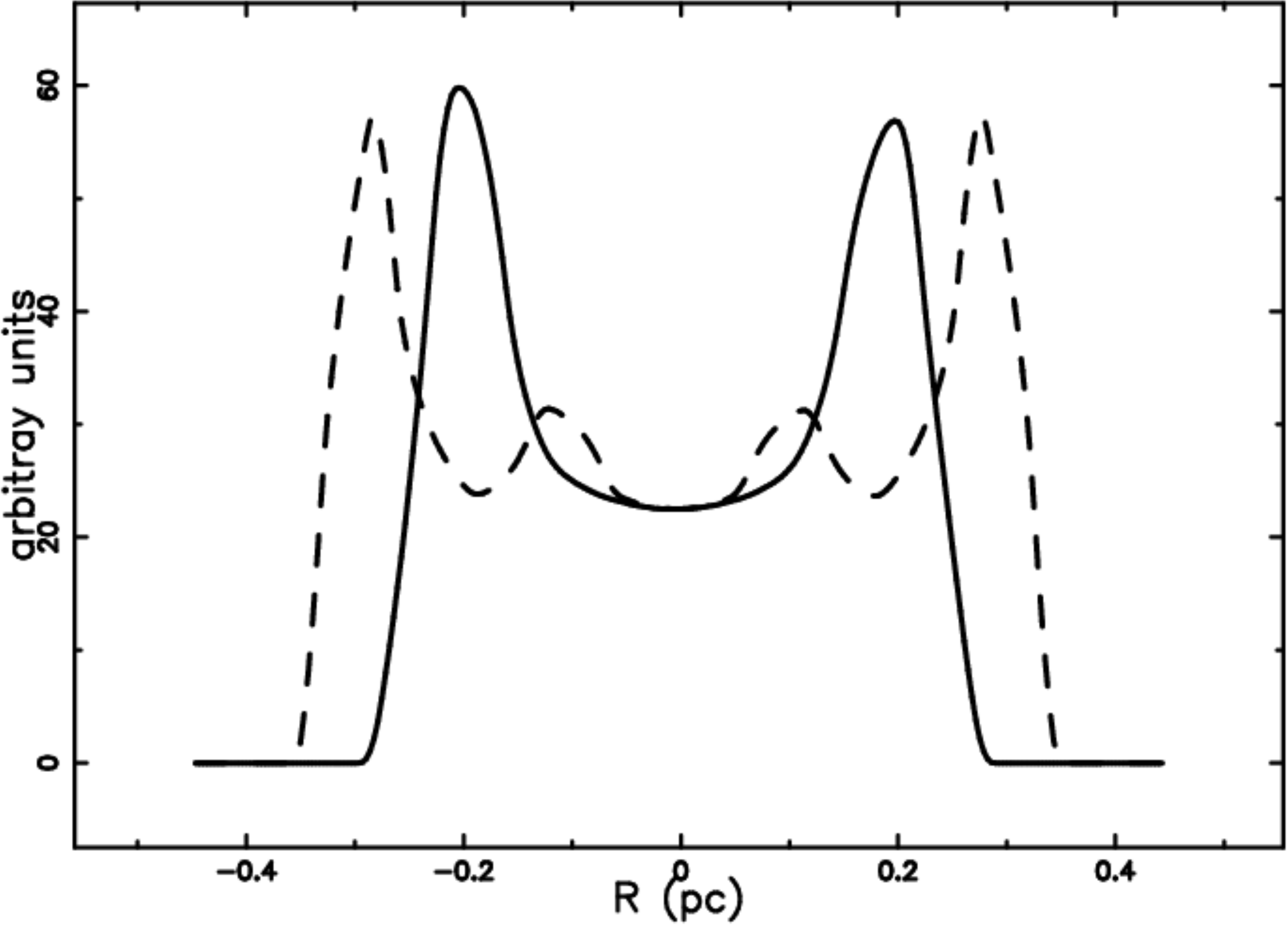}
\end {center}
\caption
{
 Two cuts of the model intensity
 across the center
 of the realistically rotated
 \sn1987a : 
 equatorial cut (full line)
 and polar cut  (dotted line).
}
\label{cut_xy_sn1987a_2}
    \end{figure*}

Two cuts of intensity 
are  reported in Figure \ref{cut_xy_sn1987a_2},
and is interesting to note that the polar cut is characterized 
by four spikes.
Here conversely the two outer lobes are explained by the image theory 
applied to a thin advancing shell.

\section{Conclusions}
{\bf Law of motion}
The presence  of a thin
self-gravitating disk of gas which is characterized by a
Maxwellian distribution
in velocity and  distribution of density which varies
only in the z-direction    produces
an axial  symmetry in the temporal evolution
of a SN.
The resulting Eq. (\ref{fundamental})
has a  form which can be solved numerically.
The  results   depend from the chosen latitude
 and more precisely  the radius as function of the time
is maximum at the poles and minimum at the equator.
A comparison
between the observed radius and the velocity at the equator,
see
\cite{Chiad2012},
gives an agreement
of $92\%$ for the theoretical versus  observed
radius and of $62\%$ for  the theoretical versus observed
velocity.
This is the {\it first} result that confirms the rationality 
of the method here presented.
The  momentum carried  away  by the
photons can be modeled  by introducing a
reasonable form for the  photon's losses , see eqn.  (
\ref{momentumlosses}).

{\bf Images}

The combination of different processes such as
the bipolar 3D shape ,
the thin layer approximation
(which means advancing layer having thickness
$\approx 1/10$ of the momentary radius),
the conversion of the rate of kinetic energy into luminosity
and the observer's point of view allows
to simulate the observed triple ring  morphology of
\sn1987a .
The framework  here adopted is similar
to that adopted to explain the
asymmetric shape  of the nebula around
$\eta$-Carinae (Homunculus),
see \cite{Zaninetti2010h,Zaninetti2012h}.
We therefore suggest that the astronomers
should map the non rotated
radius  and velocity of
\sn1987a following the outlines
of Table 1 in~\cite{Smith2006}.
These maps , if realized ,
will produce  a more clear
framework for the observed triple ring.
The suggested similarity of the triple ring system 
of  \sn1987a with  $\eta$-Carinae
is the {\it second} result which ensures 
the rationality of the method  here presented.

\noindent
{\bf REFERENCES}


\begin{thebibliography}{10}
\expandafter\ifx\csname url\endcsname\relax
  \def\url#1{{\tt #1}}\fi
\expandafter\ifx\csname urlprefix\endcsname\relax\def\urlprefix{URL }\fi
\providecommand{\eprint}[2][]{\url{#2}}

\bibitem{Chevalier1982a}
{Chevalier} R~A 1982 {Self-similar solutions for the interaction of stellar
  ejecta with an external medium} {\em \apj\/} {\bf 258}, 790

\bibitem{Chevalier1982b}
{Chevalier} R~A 1982 {The radio and X-ray emission from type II supernovae}
  {\em \apj\/} {\bf 259}, 302

\bibitem{Reynolds1986}
{Reynolds} S~P and {Gilmore} D~M 1986 {Radio observations of the remnant of the
  supernova of A.D. 1006. I - Total intensity observations} {\em \aj\/} {\bf
  92}, 1138

\bibitem{Morris2007}
{Morris} T and {Podsiadlowski} P 2007 {The Triple-Ring Nebula Around SN 1987A:
  Fingerprint of a Binary Merger} {\em Science\/} {\bf 315}, 1103
  (\textit{Preprint} \eprint{arXiv:astro-ph/0703317})

\bibitem{Tziamtzis2011}
{Tziamtzis} A, {Lundqvist} P, {Gr{\"o}ningsson} P and {Nasoudi-Shoar} S 2011
  {The outer rings of SN 1987A} {\em \aap\/} {\bf 527} A35 (\textit{Preprint}
  \eprint{1008.3387})

\bibitem{Zaninetti2012b}
{Zaninetti} L 2012 {On the spherical-axial transition in supernova remnants}
  {\em Astrophysics and Space Science\/} {\bf 337}, 581 (\textit{Preprint}
  \eprint{1109.4012})

\bibitem{Bertin2000}
{Bertin} G 2000 {\em {Dynamics of Galaxies}\/} (Cambridge: {Cambridge
  University Press.})

\bibitem{Padmanabhan_III_2002}
{Padmanabhan} P 2002 {\em {Theoretical astrophysics. Vol. III: Galaxies and
  Cosmology}\/} ({Cambridge, MA}: {Cambridge University Press})

\bibitem{Hill1828}
Hill C~J 1828 {Ueber die Integration logarithmisch-rationaler Differentiale.}
  {\em J. Reine Angew. Math.\/} {\bf 3}, 101

\bibitem{Lewin1981}
Lewin L 1981 {\em {Polylogarithms and associated functions.}\/} (New York:
  {North Holland})

\bibitem{NIST2010}
Olver F~W~J~e, Lozier D~W~e, Boisvert R~F~e and Clark C~W~e 2010 {\em {NIST
  handbook of mathematical functions.}\/} (Cambridge: {Cambridge University
  Press. })

\bibitem{press}
{Press} W~H, {Teukolsky} S~A, {Vetterling} W~T and {Flannery} B~P 1992 {\em
  {Numerical Recipes in FORTRAN. The Art of Scientific Computing}\/}
  (Cambridge: Cambridge University Press)

\bibitem{Racusin2009}
{Racusin} J~L, {Park} S, {Zhekov} S, {Burrows} D~N, {Garmire} G~P and {McCray}
  R 2009 {X-ray Evolution of SNR 1987A: The Radial Expansion} {\em \apj\/} {\bf
  703}, 1752 (\textit{Preprint} \eprint{0908.2097})

\bibitem{Chiad2012}
{Chiad} B~T, {Karim} L~M and {Ali} L~T 2012 {Study the Radial Expansion of SN
  1987A Using Counting Pixels Method} {\em International Journal of Astronomy
  and Astrophysics\/} {\bf 2}, 199

\bibitem{deyoung}
{De Young} D~S 2002 {\em {The physics of extragalactic radio sources}\/}
  (Chicago: University of Chicago Press)

\bibitem{Goldstein2002}
{Goldstein} H, {Poole} C and {Safko} J 2002 {\em {Classical mechanics}\/} (San
  Francisco: Addison-Wesley)

\bibitem{Zaninetti2010h}
{Zaninetti} L 2010 {On the 3D structure of the nebula around {$\eta$}-Carinae}
  {\em Advances in Space Research\/} {\bf 46}, 1341 (\textit{Preprint}
  \eprint{1010.2364})

\bibitem{Zaninetti2012h}
{Zaninetti} L 2012 {\em {\rm Interaction of planetary nebulae , Eta-Carinae and
  supernova remnants with the Interstellar Medium, In} {\it Interstellar
  Medium: New Research }, Cancellieri B. and Mamedov V Eds.\/} (NY: Nova
  Publishers) pp 91--146

\bibitem{Smith2006}
{Smith} N 2006 {The Structure of the Homunculus. I. Shape and Latitude
  Dependence from H2 and Fe II Velocity Maps of eta Carinae} {\em \apj\/} {\bf
  644}, 1151 (\textit{Preprint} \eprint{arXiv:astro-ph/0602464})

\end{thebibliography}
\providecommand{\newblock}{}

\end{document}